# Crossover from two- to three-dimensional behavior in superfluids

Norbert Schultka and Efstratios Manousakis

Department of Physics and Center for Materials Research and Technology
Florida State University, Tallahassee, Florida 32306

June 2, 1994

**Abstract**

We have studied the superfluid density $\rho_s$ on various size-lattices in the geometry $L \times L \times H$ by numerical simulation of the $x-y$ model using the Cluster Monte Carlo method. Applying the Kosterlitz-Thouless-Nelson renormalization group equations for the superfluid density we have been able to extrapolate to the $L \to \infty$ limit for a given value of $H$. In the superfluid phase we find that the superfluid density faithfully obeys the expected scaling law with $H$, using the experimental value for the critical exponent $\nu = 0.6705$. For the sizes of film thickness studied here the critical temperature $T_c$ and the coefficient $b$ entering the equation $T/(\rho_s H) \propto 1 - b(1 - T/T_c)^{1/2}$ are in agreement with the expected $H$-dependence deduced from general scaling ideas.

## 1 Introduction

Liquid $^4He$ is an ideal experimental testing ground for the theory of phase transitions and the related finite-size scaling (FSS) theory. Relevant physical quantities such as the specific heat $c$ or the superfluid density $\rho_s$ can be measured to a very high accuracy [1]-[5] and they can be used to check the FSS theory (see e.g. [6]). This theory was developed in order to account for the influence of the finite extent of systems which are confined in a finite geometry (e.g. a film) and at temperatures close to the critical temperature. The theory is based on the scaling hypothesis and on that finite-size effects can be observed when the bulk correlation length $\xi$ becomes of the order of the relevant size of the system (e.g., in a film the relevant size is the film thickness $H$). More precisely, the finite-size scaling hypothesis states that a dimensionless physical quantity (or the ratio of two physical quantities of the same dimensions), sufficiently close to the critical point, is a function only of the ratio $H/\xi$. For a physical quantity $O$ this simple but non-trivial statement can be expressed as follows [7]:

$$\frac{O(H,t)}{O(H=\infty,t)} = f\left(\frac{H}{\xi(H=\infty,t)}\right), \qquad (1)$$

$t$ is the reduced temperature and $f$ is a universal function. So far the validity of this approach has been confirmed by experiments on superfluid helium on Helium films of finite thickness [2] (the relevant size is the thickness). However, recent measurements of the superfluid density by Rhee, Gasparini, and Bishop [3] of Helium films seem to be in contradiction to the FSS-theory.

The singular behavior in the thermodynamic functions of liquid $^4He$ close to the superfluid transition can be understood in terms of a complex order parameter $\psi(\vec{r})$ which is the ensemble average of the helium atom boson creation operator. This ensemble average is defined inside a volume of a size much greater than the interatomic distance but much smaller than the temperature-dependent coherence length. In order to describe the physics at longer length scales, which is important very close to the critical point, we need to consider spatial fluctuations of the order parameter. These fluctuations can be taken into account by assigning a Landau-Ginzburg free energy functional $\mathcal{H}(\psi(\vec{r}))$ to each configuration of $\psi(\vec{r})$ and performing the sum of $e^{-\mathcal{H}/k_B T}$ over such configurations. The power laws governing the long distance behavior of the correlation functions and the critical exponents associated with the singular behavior of the thermodynamic quantities close to the critical point are insensitive to the precise functional form of $\mathcal{H}[\psi]$, and they are the same for an entire class of such functionals. The planar $x - y$ model belongs to the class of such Landau-Ginzburg free-energy functionals[8], and thus can be used to describe the fluctuations of the complex order parameter. In the pseudospin notation the $x - y$ model is expressed as

$$\mathcal{H} = -J \sum_{\langle i,j \rangle} \vec{s}_i \cdot \vec{s}_j, \qquad (2)$$

where the summation is over all nearest neighbors, $\vec{s} = (\cos\theta, \sin\theta)$ is a two-component vector which is constrained to be on the unit circle. The angle $\theta$ corresponds to the phase of the order parameter $\psi(\vec{r})$.

In this paper we investigate the $x - y$ model in a film geometry, i.e. planar dimensions $L$ with $L \to \infty$ and a finite thickness $H$. This study will allow us to examine directly the validity of the FSS theory.

There has been analytical and numerical work on the pure 3D $x - y$ model. Results of high-temperature-series studies can be found in [9], Monte-Carlo simulations were reported in [10]-[13], a renormalization group approach based on vortex lines was reported in [14]. In [15] the anisotropic 3D $x - y$ model ($J_x = J_y \neq J_z$) was studied. A crossover from 3D to 2D behavior was found with respect to the ratio $J_z/J_x$. The Villain model, which is in the same universality class as the $x - y$ model, was studied in a film geometry in [16] where the correlation length in the disordered phase was used to extract the thickness-dependent critical temperature.

In this paper we study the superfluid density or helicity modulus of the $x - y$ model in a film geometry, i.e. on an $L^2 \times H$ lattice. In a film geometry this model exhibits a crossover from 3D to 2D behavior. In the temperature range where the model behaves effectively two-dimensionally we are able to compute the values for the helicity modulus in the $L \to \infty$ limit using the Kosterlitz-Thouless-Nelson renormalization group equations. This enables us to eliminate the $L$-dependence completely and thus to check scaling of the helicity modulus with respect to the film thickness $H$. We also test further consequences of the FSS theory which are described in the next section. In section 3 we show how the helicity modulus is computed in our model and briefly describe the Monte-Carlo method. Section 4 discusses the $L$-dependence of the helicity modulus in the temperature range where the model exhibits two-dimensional behavior. In that section we also describe the extrapolation procedure to the $L = \infty$ limit. In section 5 we study the scaling of the helicity modulus with respect to $H$ and in section 6 we investigate the $H$-dependence of the critical temperature. The last section summarizes our results.

## 2 The $x - y$ model and finite-size scaling

The 3D $x - y$ model shows an order-disorder phase transition. Above the bulk critical temperature $T_\lambda$, the spin-spin correlation function decays exponentially. The correlation length characterizing this decay grows with the temperature $T$ according to

$$\xi(T) \propto \left(\frac{T - T_\lambda}{T_\lambda}\right)^{-\nu}. \qquad (3)$$

In the ordered phase the correlation function decays according to a power law. There, the role of a relevant finite length-scale is played by the transverse correlation length[17], $\xi_T$. $\xi_T$, up to a constant factor, is proportial to the ratio $T/\Upsilon$, where $\Upsilon$ is the helicity modulus. Fisher's scaling hypothesis implies that

$$\xi_T(T) \propto \frac{T}{\Upsilon} \propto \left(\frac{T_\lambda - T}{T_\lambda}\right)^{-\nu}. \tag{4}$$

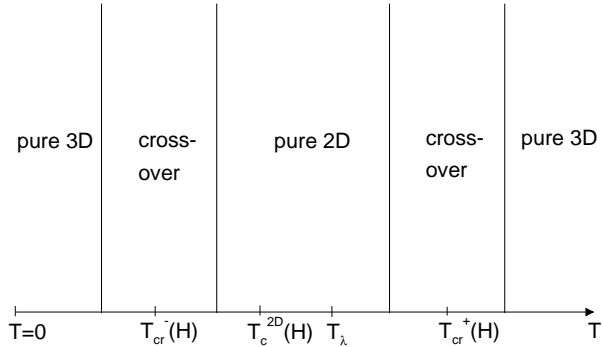

Figure 1: The behavior of the 3D $x-y$ model in a film geometry for a fixed thickness $H$ with respect to the temperature.

If the model is considered in a film geometry, i.e. infinite planar dimensions and a finite thickness $H$, interesting crossover phenomena take place. In Fig.1 we show an intuitive picture of the behavior of our model with respect to the temperature for a fixed thickness $H$. Let us start at a temperature far below $T_\lambda$ where $\xi_T$ is much smaller than the thickness $H$. At these temperatures the model behaves as a true 3D system. If we raise the temperature in order to approach closer to $T_\lambda$ from below, the correlation length $\xi_T$ grows according to (4) until we reach the crossover temperature $T_{cr}^- < T_\lambda$ where it becomes comparable to $H$. Above this temperaure the behavior of the system crosses over from the 3D to two-dimensional (2D) behavior. A further increase in temperature makes the 2D behavior of the system more and more pronounced. The system starts "feeling" a 2D critical temperature $T_c^{2D}$. Very close to $T_c^{2D}$, i.e. in the purely 2D regime, we have to apply the Kosterlitz-Thouless-Nelson theory in order to explain the behavior of the system. This implies that in this regime the correlation length $\xi_T$ does not depend on $T$ according to Eq. (4). Instead, the dimensionless ratio $K = T/(\Upsilon H)$ satisfies the following renormalization group equations [18]:

$$\frac{dK(l,T)}{dl} = 4\pi^3 y^2(l,T), \tag{5}$$

$$\frac{dy(l,T)}{dl} = (2 - \pi K^{-1}(l,T))y(l,T). \tag{6}$$

$\ln y$ is the chemical potential to create a single vortex, $e^l$ denotes the size of the core radius of a vortex. In the limit $l = \infty$, i.e. all vortices have been integrated out, and $T \to T_c^{2D}$ one finds [18]

$$K(l = \infty, T \to T_c^{2D}) = \frac{\pi}{2}\left[1 - b\left(1 - \frac{T}{T_c^{2D}}\right)^{1/2}\right], \tag{7}$$

where $b$ is a constant. $K(l = \infty, T)$ is infinite above $T_c^{2D}$, i.e. $K(l = \infty)$ exhibits a universal jump at $T_c^{2D}$. Above $T_c^{2D}$ but still in the purely 2D region, the correlation length grows with $T$ according to the Kosterlitz-Thouless theory, i.e.

$$\xi_T(T) \propto \exp B/(T - T_c^{2D})^{1/2}, \qquad (8)$$

A further increase in temperature results in reaching another crossover temperature $T_{cr}^+$, where the correlation length $\xi_T$ is comparable to the thickness $H$. At higher temperatures the model exhibits pure 3D behavior again. From this intuitive picture we deduce the following inequality for the crossover temperatures, $T_c^{2D}$, and $T_\lambda$:

$$T_{cr}^- < T_c^{2D} < T_\lambda < T_{cr}^+. \qquad (9)$$

Within the FSS theory the behavior of our model in the crossover region at temperatures $T \leq T_{cr}^-$ can be described by a universal scaling function $\Phi$ which depends only on the ratio $H/\xi_T(T)$, provided $H$ is large enough. According to Ambegaokar et al. [19] we have

$$K(T, H) = \frac{T}{\Upsilon(T, H)H} = \Phi(tH^{1/\nu}), \qquad (10)$$

where $t = (T_\lambda - T)/T_\lambda$. In the argument of $\Phi$ we have replaced $\xi_T(T)$ by its bulk scaling expression (4).

Interesting conclusions can be drawn if we extend the validity of Eq. (10) up to $T_c^{2D}$ [19]. Since $K(T,H)$ drops discontinuously to zero at $T_c^{2D}$, the scaling function $\Phi$ has to be discontinuous too. From the Kosterlitz-Thouless-Nelson theory we have to require

$$\begin{aligned} \Phi(x < x_c) &= \infty, \\ \Phi(x_c) &= \frac{\pi}{2}. \end{aligned} \qquad (11)$$

$x_c$ is a dimensionless number. From here we immediately derive an expression for the $H$-dependence of $T_c^{2D}$, namely [19]

$$T_c^{2D}(H) = T_\lambda \left(1 - \frac{x_c}{H^{1/\nu}}\right). \qquad (12)$$

Eqs. (7) and (10) have to be reconciled in the two-dimensional region. This enables us to deduce the form of the universal function $\Phi(x)$ for values of $x$ close to $x_c$. In order to do that we replace $T_c^{2D}$ by $T_\lambda$ in Eq. (7) by inverting Eq. (12). Keeping only terms linear in $H^{-1/\nu}$ we obtain

$$K(T, H) = \frac{\pi}{2} \left[1 - \frac{b}{H^{1/2\nu}} \left(tH^{1/\nu} + tx_c - x_c - \frac{x_c^2}{H^{1/\nu}}\right)^{1/2}\right]. \qquad (13)$$

For $T$ close to $T_c^{2D}$ (i.e. $t$ close to $|1 - T_c^{2D}/T_\lambda|$) we have

$$t \sim \frac{x_c}{H^{1/\nu}}, \qquad (14)$$

which yields

$$K(T, H) = \frac{\pi}{2} \left[1 - \frac{b}{H^{1/2\nu}} \left(tH^{1/\nu} - x_c\right)^{1/2}\right]. \qquad (15)$$

$K(T, H)$ can only be a universal function of $tH^{1/\nu}$ if $b$ scales according to

$$b(H) = AH^{1/2\nu}, \qquad (16)$$

i.e. we obtain

$$\Phi(x) = \frac{\pi}{2}\left(1 - A\left(x - x_c\right)^{1/2}\right),\qquad(17)$$

$A$ is a constant which will be found numerically (see later in the text). The form (17) of the universal function is valid at temperatures very close to $T_c^{2D}$, since Eq. (7) is only an approximation itself. Petschek [20] uses a different argument in order to derive (16).

## 3   The helicity modulus and Monte-Carlo method

The helicity modulus $\Upsilon$ was introduced by Fisher, Barber and Jasnow [21]. It is related to the superfluid density via [21]

$$\rho_s(T) = \left(\frac{m}{\hbar}\right)^2 \Upsilon(T). \qquad (18)$$

For the 3D $x-y$ model on a cubic lattice the definition of the helicity modulus is [10, 22]:

$$\begin{aligned}\frac{\Upsilon_\mu}{J} &= \frac{1}{V}\left\langle \sum_{\langle i,j\rangle} \cos(\theta_i - \theta_j)(\vec{e}_\mu \cdot \vec{e}_{ij})^2 \right\rangle \\ &\quad - \frac{\beta}{V}\left\langle \left(\sum_{\langle i,j\rangle}\sin(\theta_i - \theta_j)\vec{e}_\mu \cdot \vec{e}_{ij}\right)^2 \right\rangle,\end{aligned}\qquad(19)$$

where $V$ is the volume of the lattice, $\vec{e}_\mu$ is the unit vector in the corresponding bond direction, and $\vec{e}_{ij}$ is the vector connecting the lattice sites $i$ and $j$. In the following we will omit the vector index since we will always refer to the $x$-component of the helicity modulus. Note that, because of isotropy, we have $\Upsilon_x = \Upsilon_y$. The above thermal averages denoted by the angular brackets are computed according to

$$\langle O\rangle = Z^{-1}\int \prod_i d\theta_i\, O[\theta]\exp(-\beta\mathcal{H}). \qquad (20)$$

$O[\theta]$ denotes the dependence of the physical observable $O$ on the configuration $\{\theta_i\}$, the partition function $Z$ is given by

$$Z = \int \prod_i d\theta_i\, \exp(-\beta\mathcal{H}), \qquad (21)$$

where $\beta = 1/k_BT$. The expectation values (20) are computed by means of the Monte-Carlo method using Wolffs 1-cluster algorithm [23]. This algorithm successfully tackles the problem of critical slowing down [24]. We computed the helicity modulus on lattices of different sizes $L^2 \times H$, where $H = 3, 4, 6, 8, 10$ and $L = 40, 60, 100$ for each thickness. For some thicknesses we used $L = 25, 50, 100$. Periodic boundary conditions were applied in all directions. We carried out of the order of $10,000$ thermalization steps and of the order of $500,000$ measurements. The calculations were performed on a heterogenes environment of workstations which include Sun, IBM RS/6000 and DEC alpha AXP workstations and on the Cray-YMP.

# 4 The two-dimensional region.

Here we consider the temperature range $T_{cr}^- < T < T_{cr}^+$, where the 3D bulk correlation length exceeds the thickness of the film. This temperature range contains both the $H$-dependent 2D critical temperature $T_c^{2D}$ as well as the 3D bulk critical temperature $T_\lambda$. For fixed $H$ and for temperatures close enough to $T_c^{2D}$ the system behaves as a 2D $x-y$ model, thus, our method of analysis applied to the 2D $x-y$ model [25] can be used here as well. In [25] we investigated the dimensionless quantitiy $\Upsilon/J$, in the 3D system the helicity modulus aquires the additional unit of a length$^{-1}$, thus the proper quantity to consider is the ratio $\Upsilon(L,H,T)H/J$.

In the following sections we always leave $H$ fixed.

## 4.1 Finite-size scaling with respect to $L$ above $T_c^{2D}(H)$.

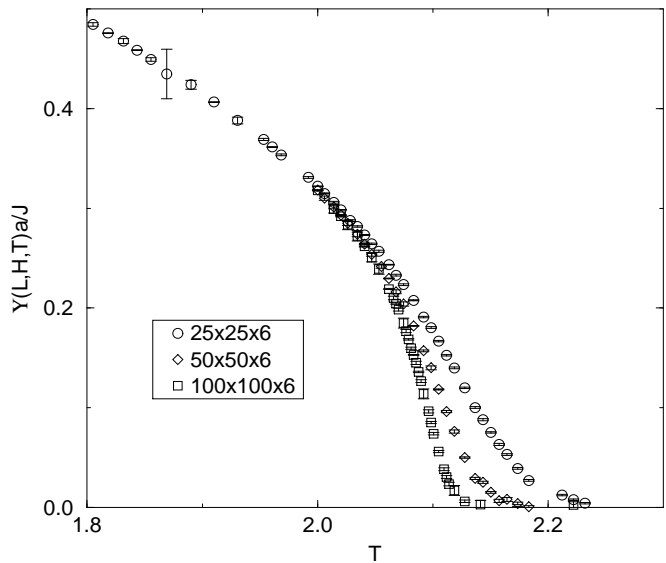

Figure 2: The helicity modulus $\Upsilon(L, H, T)$ as a function of $T$ for various lattices $L^2 \times 6$.

In Fig.2 we show the data for $\Upsilon(L, H, T)a/J$ for films of fixed thickness $H = 6$ for various sizes $L$, $a$ denotes the lattice spacing. As in [25] we can obtain a function $T = F(L)$ via the beta function such, that $\Upsilon(L', H, F(L')) = \Upsilon(L, H, F(L)) = \Upsilon_p$, where $\Upsilon_p$ is a physical value of the helicity modulus. We define the beta function as:

$$\beta(T) = -\lim_{L \to \infty} \frac{dT}{d\ln(L)}. \qquad (22)$$

Since we expect our model to behave effectively two-dimensionally we use the ansatz

$$\begin{aligned} \beta(T > T_c(H)) &= c(H)(T - T_c^{2D}(H))^{1+\nu}, \\ \beta(T \leq T_c(H)) &= 0, \end{aligned} \qquad (23)$$

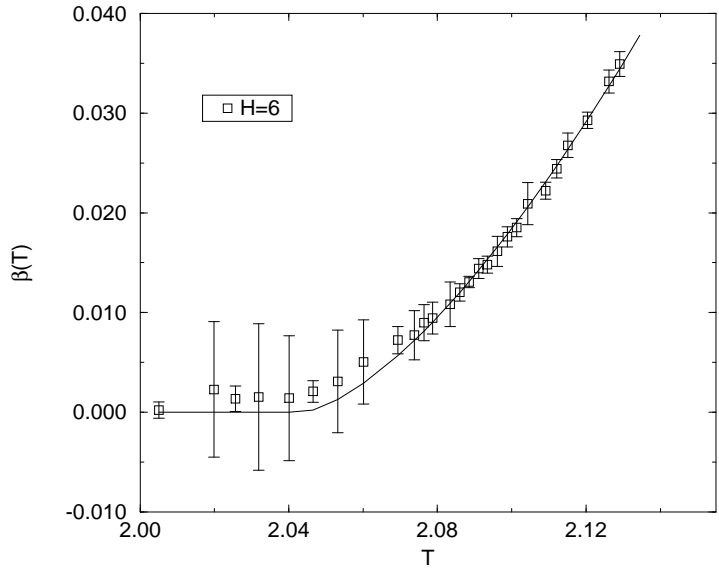

Figure 3: The beta function obtained from the $100^2 \times 6$ and $50^2 \times 6$ lattices.

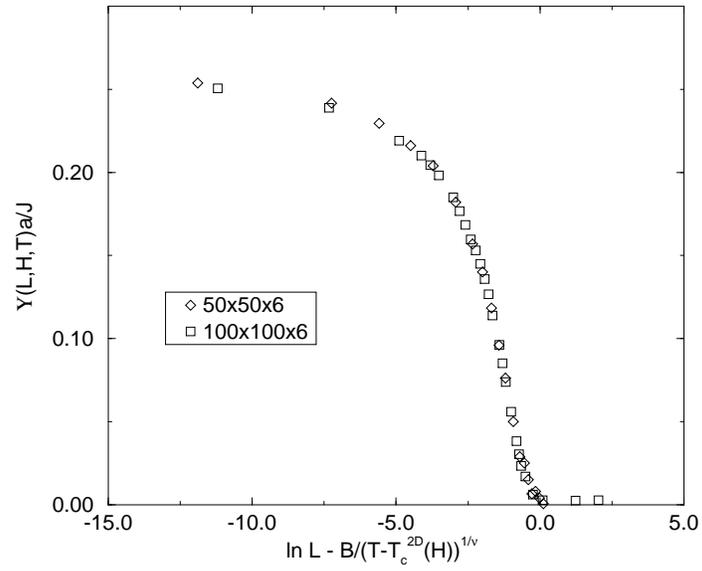

Figure 4: The helicity modulus $\Upsilon$ as a function of $z$ for the $100^2 \times 6$ and $50^2 \times 6$ lattices.

which is suggested by the Kosterlitz-Thouless theory. Inserting the ansatz (23) into (22) and integrating yields:

$$\ln L - \frac{B}{(T - T_c(H))^\nu} = z, \qquad (24)$$

where $B = 1/(\nu c)$ and $z$ is a constant of integration, which depends on the value of $\Upsilon$ used to define the scaling transformation. As in the pure 2D case we expect all values of $\Upsilon(L, H, T)/J$ for fixed $H$ to collapse on the same universal curve if $\Upsilon/J$ is considered as a function of $z$. As was explained in [25] this also means, that the correlation length $\xi$ grows according to $\xi(T) \propto \exp B/(T - T_c(H))^\nu$ because $\Upsilon/J$ for fixed $H$ should be a function of $L/\xi(T)$ only, and thus, also a function of $z = \ln(L/\xi(T))$. The beta function can be found numerically using methods described in [25]-[28].

Since the critical temperature $T_c^{2D}$, which enters in the expression for the beta function, is very sensitive to the planar extension $L$ of the lattice [25] in contrast to the other parameters $c$ and $\nu$, we only use the two largest lattices to numerically derive the beta function for each different thickness $H$. This way we determine a lower bound for the critical temperature $T_c^{2D}(H)$. Having extracted data points for the beta funtion we fit them to the functional form (23) setting $\nu = 0.5$. The results of our fits are given in Table 1. Fig.3 shows the beta function and Fig.4 demonstrates that the values for the quantity $\Upsilon a/J$ for the $50^2 \times 6$ and $100^2 \times 6$ lattices collapse on one curve.

| $H$ | $L_1, L_2$ | $c$ | $T_c^{2D}(H)$ | $\chi^2$ | $Q$ |
|---|---|---|---|---|---|
| 3 | 50,100 | 0.946(29) | 1.7710(23) | 0.57 | 0.92 |
| 4 | 60,100 | 0.955(41) | 1.9039(26) | 0.66 | 0.92 |
| 6 | 50,100 | 1.383(58) | 2.0387(17) | 0.55 | 0.96 |
| 8 | 60,100 | 1.10(10) | 2.0813(31) | 0.50 | 0.98 |
| 10 | 60,100 | 1.66(18) | 2.1182(30) | 0.29 | 1.0 |

Table 1: Fitted values of the parameters $c(H)$ and $T_c^{2D}(H)$ of the beta function (23) for the used lattice pairs for different thicknesses $H$, $\nu = 0.5$. $\chi^2$ and the goodness of the fit $Q$ are also given.

## 4.2 Finite-size scaling with respect to $L$ below $T_c^{2D}(H)$.

Since our calculation was performed on finite lattices of size $L^2 \times H$ we wish to reach the $L \to \infty$ limit and to obtain $T_c^{2D}(H)$ and $b(H)$ (see Eq. (7)). In order to do that we need to know the leading finite-$L$ corrections to the dimensionless ratio $K = T/(\Upsilon H)$. As the system behaves effectively two-dimensionally we can apply the formulae derived in [25] for the ratio $K$. This ratio satisfies the renormalization group equations (5) and (6), thus solving these equations for a finite scale $l = \ln L$ in the limit $L \to \infty$ and close to the critical temperature $T_c^{2D}(H)$ yields [25]:

$$K(L \to \infty, T < T_c(H)) = K_\infty(T) \left(1 + \frac{2(1 - 2K_\infty(T)/\pi)}{1 - \tilde{c} L^{2(\pi/K_\infty(T) - 2)}}\right), \qquad (25)$$

$$K(L, T_c(H)) = \frac{\pi}{2}\left(1 - \frac{1/2}{\ln L + \tilde{c}'}\right), \qquad (26)$$

where $K_\infty(T) \equiv K(L \to \infty, T)$ and $\tilde{c}$ and $\tilde{c}'$ are constants of integration.

At a fixed thickness $H$ below $T_c^{2D}(H)$ we can use the expression (25) to extrapolate our values for $K(L, T)$ at finite $L$ to the values $K_\infty(T)$ at infinite $L$. In order to do that we fit our calculated values for $K(L, T)$ to the functional form (25). Table 2 contains our fitting results and Fig.5 shows a typical fit. Note that the $H$-dependence of the ratio $K$ is contained in the $H$-dependence of the critical temperature only.

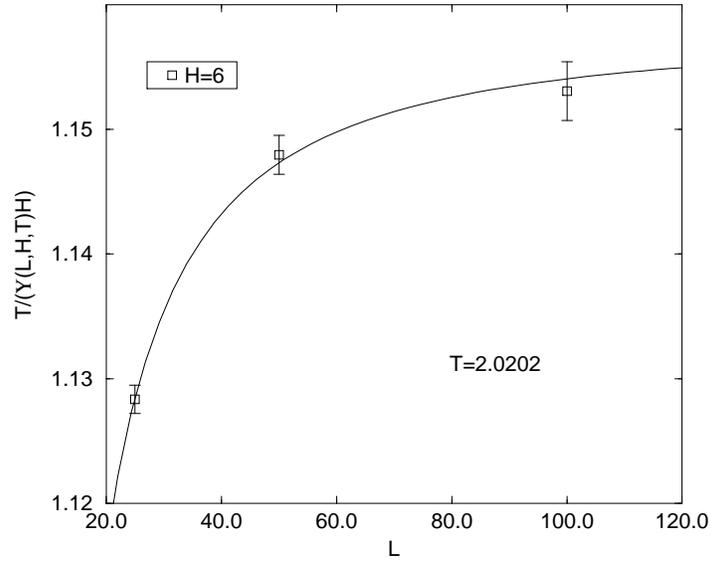

Figure 5: $T/(\Upsilon(L,H,T)H)$ as a function of $L$ at $T = 2.0202$ and $H = 6$. The solid curve is the fit to (25).

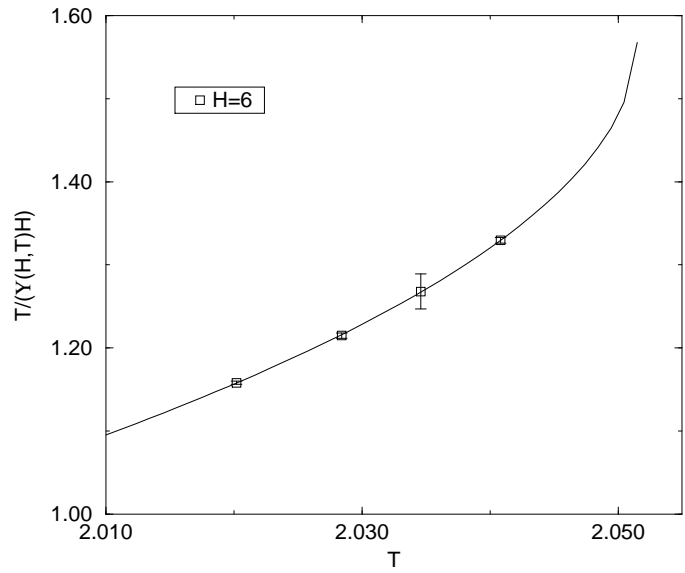

Figure 6: $T/(\Upsilon(H,T)H)$ at $L = \infty$ and $H = 6$ as a function of $T$. The solid curve is the fit to (27).

| $H$ | $T$ | $\tilde{c}$ | $K_\infty(H,T)$ |
|---|---|---|---|
| 3 | 1.7699 | 1.061(33) | 1.3363(14) |
|   | 1.7544 | 0.838(55) | 1.2691(13) |
|   | 1.7391 | 0.698(30) | 1.2150(4) |
| 4 | 1.9048 | 0.663(44) | 1.2834(20) |
|   | 1.8868 | 0.428(66) | 1.1968(20) |
|   | 1.8692 | 0.327(75) | 1.1294(9) |
| 6 | 2.0408 | 0.566(9) | 1.3296(36) |
|   | 2.0346 | 0.396(37) | 1.268(21) |
|   | 2.0284 | 0.324(32) | 1.2148(33) |
|   | 2.0202 | 0.219(12) | 1.1580(19) |
| 8 | 2.0964 | 0.433(10) | 1.338(12) |
|   | 2.0899 | 0.241(26) | 1.2352(52) |
|   | 2.0881 | 0.217(33) | 1.2104(63) |
|   | 2.0877 | 0.205(26) | 1.2054(52) |
|   | 2.0872 | 0.198(33) | 1.1999(67) |
| 10 | 2.1277 | 0.518(21) | 1.404(12) |
|    | 2.1266 | 0.436(14) | 1.359(15) |
|    | 2.12653 | 0.432(14) | 1.358(15) |
|    | 2.12648 | 0.429(14) | 1.356(14) |
|    | 2.12544 | 0.356(17) | 1.318(13) |
|    | 2.12540 | 0.353(17) | 1.317(13) |
|    | 2.12535 | 0.351(16) | 1.317(12) |

Table 2: Fitted values of the parameters $\tilde{c}$ and $K_\infty(H,T)$ of the expression (25) for various temperatures $T$ and different thicknesses $H$.

The extrapolated values for $K_\infty(T)$ at a fixed $H$ should behave as [18]:

$$K_\infty(T \to T_c^{2D}(H)) = \frac{\pi}{2}\left[1 - b(H)\left(1 - \frac{T}{T_c^{2D}(H)}\right)^{1/2}\right], \qquad (27)$$

where we introduced the expected $H$-dependence of the parameters $b$ and $T_c^{2D}$. At a fixed $H$ both parameters can be determined by a fit of our results for $K_\infty(T)$ to the functional form (27). The results of our fits are presented in Table 3. In Fig.6 the fit to the data for $K_\infty(T,H)$ at $H=6$ is shown.

| $H$ | $b(H)$ | $T_c^{2D}(H)$ | $\chi^2$ | $Q$ |
|---|---|---|---|---|
| 3 | 1.3005(61) | 1.7935(5) | 0.009 | 0.92 |
| 4 | 1.5701(94) | 1.9310(6) | 0.15 | 0.70 |
| 6 | 2.128(22) | 2.0515(5) | 0.07 | 0.94 |
| 8 | 2.78(10) | 2.1023(10) | 0.03 | 0.99 |
| 10 | 3.73(28) | 2.1294(5) | 0.006 | 1.0 |

Table 3: Fitted values of the parameters $b(H)$ and $T_c^{2D}(H)$ of the expression (27) for different thicknesses $H$. $\chi^2$ and the goodness of the fit $Q$ are also given.

Note that the $H$-dependence of the ratio $K$ is contained in the $H$-dependence of the critical temperature $T_c^{2D}$ and the constant $b$. These two parameters are the only parameters which are free to depend on the film thickness if we request the form of the equations (25), (26) and (27) to be applicable to our case of a film geometry.

We would like to mention one difficulty in extracting the infinite $L$ results $K_\infty(T)$ and finding the parameters $T_c^{2D}(H)$ and $b(H)$. In our study of the pure 2D $x-y$ model [25] we found that formulae (25) and (27) are valid in a narrow interval below $T_c^{2D}$, in particular the expression (27) is a good approximation for temperatures in the region $\sim 0.9 T_c^{2D} \leq T \leq T_c^{2D}$. For films of finite thickness $H$ we have the additional problem of a crossover between 3D and 2D behavior, which makes the region of validity of expression (27) even smaller. Thus, one has to be careful in chosing the correct temperature range for the extrapolation procedure.

## 5  Scaling of $T/(\Upsilon H)$ with respect to $H$.

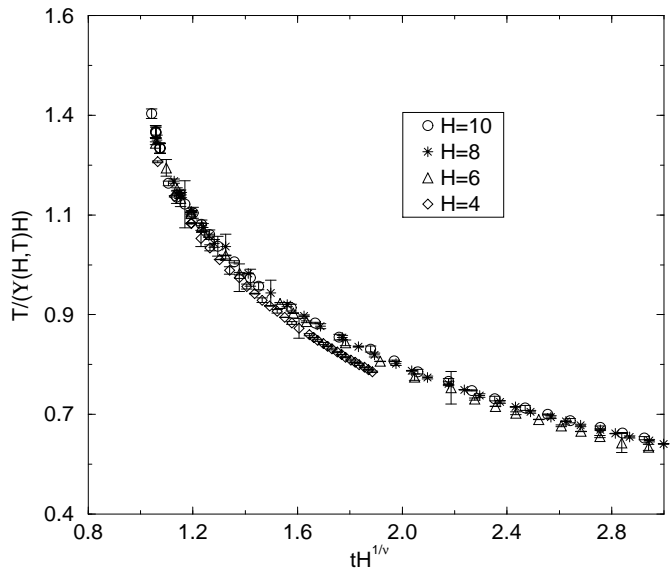

Figure 7: $T/(\Upsilon(H,T)H)$ as a function of $tH^{1/\nu}$. $\nu = 0.6705$. Only even number thicknesses are displayed.

This section is concerned with the scaling relation (10). In order to check the validity of the scaling form (10) we plot first $T/(\Upsilon H)$ versus $tH^{1/\nu}$ using the experimental value of Goldner and Ahlers $\nu = .6705$ [29] and the result of reference [13] $T_\lambda = 2.2017$. This is done in Fig.7. In the plot we only use temperatures below $T_c^{2D}(H)$. If the scaling behavior (10) is valid, all our data points should lie on one universal curve. This is not the case for the films of thickness $H = 3, 4$. For the other films scaling seems valid in the interval $1.0 \leq tH^{1/\nu} \leq 2.5$. All the represented data have lost their $L$-dependence within errorbars. Scaling is confirmed by Fig.8 which shows the same plot as Fig.7, but only the data for the thicknesses $H = 6, 8, 10$ are displayed. The data points collapse onto one universal curve in the specified interval. The films with thickness $H = 3, 4$ are too thin in order to exhibit the scaling behavior (10).

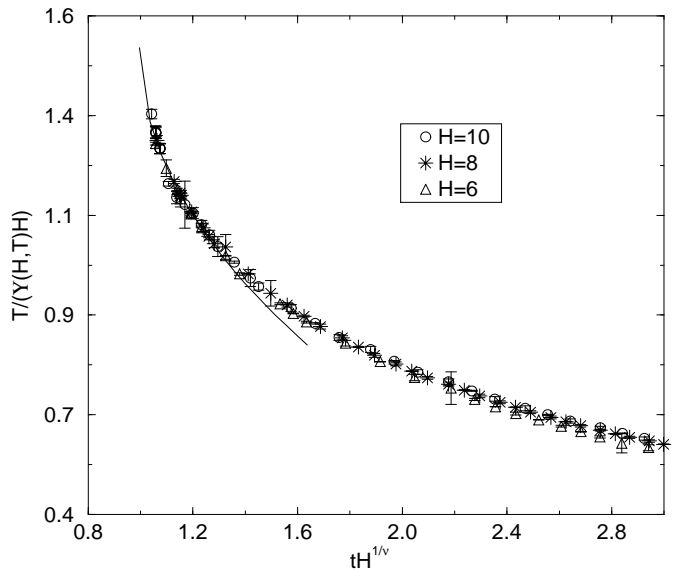

Figure 8: $T/(\Upsilon(H,T)H)$ as a function of $tH^{1/\nu}$. $\nu = 0.6705$. The solid curve is expression (17) with the parameters (28).

We would like to compare our findings to the experimental results for the superfluid density of very thick films by Rhee, Gasparini and Bishop (RGB) [3]. In our language they plot $\Upsilon H$ versus $tH^{1/\nu}$ and find that their data do not collapse for the expected value of $\nu$. RGB demonstrated the lack of scaling of their data by collapsing them on one universal curve using a different value of $\nu$ ($\nu = 1.14(02)$). Since we confirm scaling for the ratio $K(T, H)$ in our simulation the lack of scaling of the experimental data is not due to a breakdown of the phenomenological scaling theory. Note that our results have been obtained with periodic boundary conditions. We are in the process of repeating the calculation using Dirichlet boundary conditions which we believe represent the experimental situation more closely.

Since the scaling function $\Phi(x)$ is known for temperatures which correspond to values of $x$ close to $x_c$ we can actually find the constants $A$ and $x_c$ in expression (17). For this purpose we plot $K_\infty(H,T)$ given in Table 2 versus $tH^{1/\nu}$ with the values for $T_\lambda$ and $\nu$ as above for the three thickest films and fit the resulting data points to the form (17). We obtain

$$\begin{aligned} A &= 0.5928 \pm 0.0050, \\ x_c &= 0.9964 \pm 0.0023. \end{aligned} \qquad (28)$$

The fitted curve is the solid line in Fig.8. The universal function (17) with the parameters (28) describes the collapsed data rather well in the interval $1.0 \leq tH^{1/\nu} \leq 1.3$. Since $T/\Upsilon(T) \propto \xi_T(T)$ we would like to point out that $\Phi(1.3) \sim 1$, i.e. $\xi_T \sim H$. This agrees with the general picture that the 2D behavior sets in when the 3D bulk correlation length exceeds the film thickness.

# 6  $H$-dependence of fitting parameters.

In this section we would like to examine the $H$-dependence of $T_c^{2D}$ and $b$.

| data points | $x_c T_\lambda$ | $1/\nu$ | $T_\lambda$ | $\chi^2$ | $Q$ |
|---|---|---|---|---|---|
| 5 | 2.006(21) | 1.451(12) | 2.2008(17) | 4.75 | 0.0087 |
| 4 | 2.177(64) | 1.523(27) | 2.1947(25) | 0.39 | 0.53 |

Table 4: Fitted values of the parameters $x_c T_\lambda$, $1/\nu$, and $T_\lambda$ of the expression (12) for various numbers of datapoints included in the fit. $\chi^2$ and the goodness of the fit $Q$ are also given.

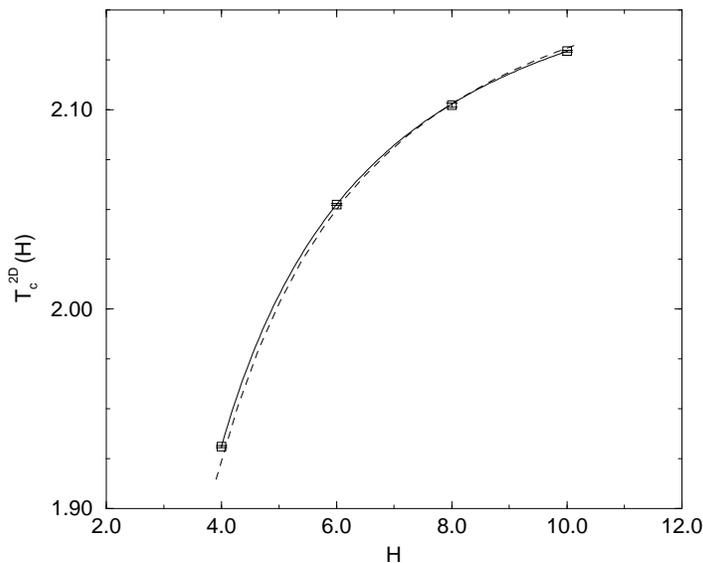

Figure 9: $T_c^{2D}(H)$ as a function of $H$. The solid curve is the fit to (12), the dashed line represents (29).

We fit our results for $T_c^{2D}(H)$ (see Table 3) to the expression (12) as expected. The results of the fit to the form (12) for all critical temperatures and the critical temperatures corresponding to the four thickest films are collected in Table 4, respectively. The fit including the critical temperatures of the four thickest films is shown in Fig.9. The bulk critical temperature $T_\lambda$ obtained from the fit (which uses five data points) agrees within error bars with $T_\lambda = 2.2017 \pm 0.0005$ obtained by Janke [13] and Hasenbusch and Meyer [12]. The critical exponent turns out to be $\nu = 0.6892 \pm 0.0057$ which is a little larger than the expected value. The same fit to only four critical temperatures yields $\nu = 0.657 \pm 0.012$ and $T_\lambda = 2.1947 \pm 0.0025$. The critical exponent agrees with the expected value within errorbars, but $T_\lambda$ is somewhat smaller than Janke's value. The critical exponent $\nu$ was estimated experimentally as well. In experiments on bulk helium Goldner and Ahlers [29] and Singsaas and Ahlers [30] deduced $\nu = 0.6705 \pm 0.0006$ and $\nu = 0.6717 \pm 0.0004$, respectively.

We can obtain the same curves as described above using the parameters (28) entering the scaling expres-

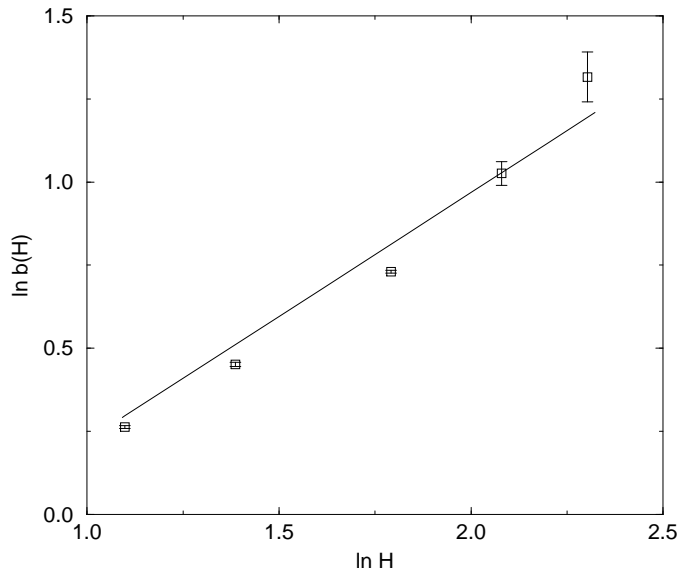

Figure 10: $\ln b(H)$ as a function of $\ln H$. The solid curve represents (30).

sions (12) and (16). The dashed line in Fig.9 represents the function

$$T_c^{2D}(H) = 2.2017 \left(1.0 - \frac{0.9964}{H^{1./0.6705}}\right). \tag{29}$$

This curve is in fairly good agreement with the critical temperatures for the three thickest films, which we should expect because of scaling for these films. In Fig.10 we plot $b$ versus $H$ and the function

$$b(H) = 0.5928 H^{0.5/0.6705}. \tag{30}$$

Here the agreement for the three thickest films is not as good as for the case of the critical temperatures $T_c^{2D}(H)$. However we attribute the scattering of the data points around the curve (30) to the procedure we applied in order to extract $b(H)$ and $T_c^{2D}(H)$. A possible cause of the poor agreement may be that more data points in the 2D critical region are required. Furthermore we have always assumed in the fitting procedure that the square root form in the expressions (7) and (17) is correct. A general exponent $\mu$ leads to the following scaling form for $b$:

$$b(H) = AH^{\mu/\nu}. \tag{31}$$

A slight deviation of $\mu$ from 0.5 could also account for a better agreement of the curve (30) with the results for $b(H)$.

In summary we can state that our results for the critical temperatures $T_c^{2D}(H)$ and for the parameter $b(H)$ are in agreement with the expected $H$-dependence, which is deduced from general scaling ideas.

# 7 Summary

We have investigated the finite-size scaling properties of the superfluid density of a superfluid with respect to the film thickness. This was done by means of a Monte-Carlo simulation of the $x - y$ model in a $L \times L \times H$ geometry with periodic boundary conditions in all directions. We extrapolated the values of the superfluid density to the $L \to \infty$ limit in the critical region where the model is effectively two-dimensional using the Kosterlitz-Thouless-Nelson renormalization group equations (5) and (6). The test of the scaling expression (10) revealed that scaling for the quantity $T/(\Upsilon(T, H)H)$ is fulfilled, i.e. the numerical results for this ratio collapse on one universal curve for sizes of film thickness $H = 6, 8, 10$ used in our Monte-Carlo simulation. Furthermore we derived an analytic expression for the universal curve, which is valid for temperatures close to the critical temperatures of each film. Using the expression (7) we were able to extract the critical temperatures $T_c^{2D}(H)$ and the parameters $b(H)$ entering Eq. (27). The $H$-dependence of these two parameters agrees with the expected behavior (12) and (16) deduced from general scaling arguments.

# 8 Acknowledgements

This work was supported by the National Aeronautics and Space Administration under grant no. NAGW-3326.

# References

[1] D. J. Bishop, J. D. Reppy, Phys. Rev. Lett. **40** 1727 (1978).

[2] J. Maps, R. B. Hallock, Phys. Rev. Lett **47** 1533 (1981).

[3] I. Rhee, F. M. Gasparini, D. J. Bishop, Phys. Rev. Lett. **63** 410 (1989).

[4] D. Finotello, Y. Y. Yu, F. M. Gasparini, Phys. Rev. **B41** 10994 (1990).

[5] Y. Y. Yu, D. Finotello, F. M. Gasparini, Phys. Rev. **B39** 6519 (1989).

[6] M. E. Fisher, M. N. Barber, Phys. Rev. Lett. **28** 1516 (1972); M. E. Fisher, Rev. Mod. Phys. **46** 597 (1974); V. Privman, Finite Size Scaling and Numerical Simulation of Statistical systems, Singapore: World Scientific 1990.

[7] E. Brezin, J. Physique **43** 15 (1982).

[8] H. Kleinert, Gauge Fields in Condensed Matter, Singapore: World Scientific 1989.

[9] P. Butera, M. Comi, A. J. Guttmann, Phys. Rev. **B48** 13987 (1993); R. G. Bowers, G. S. Joyce, Phys. Rev. Lett. **19** 630 (1967); M. Ferer, M. A. Moore, M. Wortis, Phys. Rev. **B8**, 5205 (1973).

[10] Y.-H. Li, S. Teitel, Phys. Rev. **B40** 9122 (1989).

[11] G. Kohring, R. E. Shrock, P. Wills, Phys. Rev. Lett. **57** 1358 (1986); A.P. Gottlob, M. Hasenbusch, S. Meyer, Nucl. Phys. **B(Proc. Suppl.)30** 838 (1993).

[12] M. Hasenbusch, S. Meyer, Phys. Lett. **B241** 238 (1990).

[13] W. Janke, Phys. Lett. **A148** 306 (1992).

[14] G. A. Williams, Phys. Rev. Lett. **59** 1926 (1987); S. R. Shenoy, Phys. Rev. **B40** 5056 (1989).

[15] S. T. Chui, M. R. Giri, Phys. Lett. **A128** 49 (1988); W. Janke, T. Matsui, Phys. Rev. **B42** 10673 (1990).

[16] W. Janke, K. Nather, Phys. Rev. **B48** 15807 (1993).

[17] P. C. Hohenberg, A. Aharony, B.I. Halperin, E. D. Siggia, Phys. Rev. **B13** 2986 (1976).

[18] D. R. Nelson, J. M. Kosterlitz, Phys. Rev. Lett. **39** (1977) 1201.

[19] V. Ambegaokar, B. I. Halperin, D. R. Nelson and E. D. Siggia, Phys. Rev. **B21** 1806 (1980).

[20] R. G. Petschek, Phys. Rev. Lett. **57** 501 (1986).

[21] M. E. Fisher, M. N. Barber, D. Jasnov, Phys. Rev. **B16** 2032 (1977).

[22] S. Teitel, C. Jayaprakash, Phys. Rev. **B27** 598 (1983).

[23] U. Wolff, Phys. Rev. Lett. **62** 361 (1989).

[24] U. Wolff, Nucl. Phys. **B322** 759 (1989); Nucl. Phys. **B(Proc. Suppl.)**17 93 (1990).

[25] N. Schultka, E. Manousakis, Phys. Rev. **B49** 12071 (1994).

[26] P. Harten and P. Suranyi, Nucl. Phys. **B265** [FS15] 615 (1989).

[27] P. Hasenfratz, unpublished.

[28] E. Manousakis, Rev. Mod. Phys. **63** 1 (1991).

[29] L. S. Goldner, G. Ahlers, Phys. Rev. **B45** 13129 (1992).

[30] A. Singsaas, G. Ahlers, Phys. Rev **B30** 5103 (1984).